\documentclass[11pt]{article}

\usepackage{amsmath}
\usepackage{graphicx}
\usepackage{subfigure}
\usepackage{cite}
\usepackage{color}

\begin{document}

\title{\bf{Thermodynamics of Warped Anti-de Sitter Black Holes under Scattering of Scalar Field}}

\date{}
\maketitle

\begin{center}
\author{Bogeun Gwak}$^a$\footnote{rasenis@dgu.ac.kr}\\

\vskip 0.25in
$^{a}$\it{Division of Physics and Semiconductor Science, Dongguk University, Seoul 04620,\\Republic of Korea}\\
\end{center}
\vskip 0.6in

{\abstract
{We investigate the thermodynamics and stability of the horizons in warped anti-de Sitter black holes of the new massive gravity under the scattering of a massive scalar field. Under scattering, conserved quantities can be transferred from the scalar field to the black hole, which change the state of the black hole. We determine that the changes in the black hole are well coincident with the laws of thermodynamics. In particular, the Hawking temperature of the black hole cannot be zero in the process as per the third law of thermodynamics. Furthermore, the black hole cannot be overspun beyond the extremal condition under the scattering of any mode of the scalar field.}
}

\thispagestyle{empty}
\newpage
\setcounter{page}{1}
\section{Introduction}

Black holes are one of the compact objects formed by the concentration of matter in a small space. The event horizon, which is a distinct point in a black hole, is a coordinate singular surface from which light cannot escape to a static observer. Hence, the event horizon plays a role in preventing the observer from viewing the inside of a black hole. Although a particle going through the horizon cannot be seen from the outside region, its physical quantities affect the black hole through a back-reaction. Based on the above, black holes energies were divided into two parts\cite{Christodoulou:1970wf,Christodoulou:1972kt}: the irreducible mass, which always increases, even if a Penrose process extracts energy from the black hole\cite{Penrose:1971uk}, and the reducible energy, which can be extracted by a Penrose process involving rotational or electric energy. As energy, the irreducible mass is considered to be distributed on the surface area of the horizon\cite{Smarr:1972kt} and is proportional to the square-root of the horizon surface area. As its irreducible nature is similar to the second law of thermodynamics, it contributes to the Bekenstein-Hawking entropy\cite{Bekenstein:1973ur,Bekenstein:1974ax}. Furthermore, because of the quantum effects near the horizon, there is an emission toward the outside\cite{Hawking:1974sw,Hawking:1976de}; then, black holes can be regarded as thermodynamic systems with the Hawking temperature. Therefore, the laws of thermodynamics can be defined at the surface of the horizon.

The curvature singularity is located inside a black hole, and the event horizon conceals it from an outside observer. Without the horizon, the naked singularity breaks the causal structure. Hence, the horizon is conjectured to conceal the singularity from the observer, according to the weak cosmic censorship (WCC) conjecture\cite{Penrose:1964wq,Penrose:1969pc}. The horizon should be stable through perturbation to cover the inside of the black hole as per the WCC conjecture. The first test on the WCC conjecture was conducted on the Kerr black hole\cite{Wald:1974ge}. When the Kerr black hole is overspun beyond the extremal bound, the horizon disappears, and it becomes a naked singularity. However, the Kerr black hole cannot be overspun by adding a particle\cite{Wald:1974ge}. Further, various studies on the WCC conjecture are in progress, based on different points of view\cite{Jacobson:2009kt,Barausse:2010ka,Colleoni:2015afa,Gwak:2015fsa,Sorce:2017dst,Gao:2012ca,Rocha:2014jma,Cardoso:2015xtj,Siahaan:2015ljs,Gwak:2016gwj,Revelar:2017sem,Gim:2018axz,Yu:2018eqq,Zeng:2019jrh,Wang:2019jzz,He:2019kws,Hu:2019zxr,Wang:2020osg,Jiang:2020fgr,Shaymatov:2020wtj,Ying:2020bch}. The conclusion of the WCC conjecture depends on what is considered as interactions between a black hole and particle, in particular\cite{Hubeny:1998ga,Isoyama:2011ea}. Instead of a particle, the WCC conjecture can be tested using a scalar field\cite{Hod:2008zza,Semiz:2005gs,Toth:2011ab,Natario:2016bay,Duztas:2017lxk,Gwak:2018akg,Gwak:2019asi,Duztas:2019ick,Chen:2019nsr,Natario:2019iex,Jiang:2019vww,Wang:2019bml,Gwak:2019rcz,Yang:2020iat,Hong:2020zcf,Feng:2020tyc,Yang:2020czk,Gwak:2020zht,Goncalves:2020ccm}. The back-reaction of a black hole due to a particle as well as scalar field is restricted by the dispersion relationship of the matter going into the black hole. This dispersion relation can be rewritten as the first law of thermodynamics. Hence, with matter, satisfying its own dispersion relationship, it is difficult to overspin or overcharge the black hole beyond its extremal bound\cite{Gwak:2015fsa,Gwak:2018akg}. In the process, the second law of thermodynamics is also ensured. This can be demonstrated in various black holes; hence, the laws of thermodynamics can be closely related to the WCC conjecture.

Gravity theories with a negative cosmological constant the vacuum solution to the anti-de Sitter (AdS) spacetime. The gravity theory in bulk AdS spacetime corresponds to quantum theory, such as the conformal field theory (CFT) defined on the AdS boundary. This AdS/CFT correspondence is constructed in association with the $D$-dimensional gravity theory in AdS spacetime and the $D-1$-dimensional CFT\cite{Maldacena:1997re,Gubser:1998bc,Witten:1998qj,Aharony:1999ti}. Here, the pure AdS spacetime corresponds to zero-temperature CFT. CFT with finite temperature can be constructed using black holes and thermodynamics because Hawking radiation provides the temperature of the dual CFT. Hence, the Hawking temperature is the that of the CFT\cite{Witten:1998zw}. The dual description, in particular, is well studied for a rotating AdS$_3$ black hole, where there are diffeomorphism generators preserving its boundary condition. Then, Virasoro algebra can be constructed using the generators\cite{Brown:1986nw}. According the central charge of the Virasoro algebra, the dual CFT$_2$ corresponding to the AdS$_3$ black hole can be specified\cite{Strominger:1997eq}. The AdS/CFT correspondence is now extended to various applications such as quantum chromodynamics and condensed matter theory. Such correspondence is found in various non-AdS spacetimes. The warped AdS$_3$ (WAdS$_3$) black hole is a deformed AdS black hole in non-AdS spacetime\cite{Bengtsson:2005zj}. Therefore, based on the diffeomorphism generators, the WAdS$_3$ black hole is found to be associated with the warped conformal field theory in two dimensions (WCFT$_2$), which has infinite conserved charges obeying the Virasoro-Kac-Moody U(1) algebra\cite{Detournay:2012pc}. This is called the WAdS$_3$/WCFT$_2$ correspondence\cite{Anninos:2008fx,Donnay:2015iia}. Here, we consider WAdS$_3$ black holes in the new massive gravity (NMG), which is the parity-violation resolved version of the topologically massive gravity (TMG)\cite{Bergshoeff:2009hq,Donnay:2015joa}.

In this work, we investigate the thermodynamics and stability of the horizon in a WAdS$_3$ black hole, including their association with the WAdS$_3$/WCFT$_2$ correspondence. We consider the changes in the black hole under the scattering of a massive scalar field. Based on the NMG, the black hole provides interesting results under a spin-2 particle with two polarizations. Furthermore, its non-AdS boundary, WAdS$_3$, is crucial for studying the scattering in deformed AdS spacetime, and the correspondence. The carried conserved quantities under scattering cause changes in the black hole affecting its various properties. In particular, we investigate the changes in the thermodynamic variables; these variable changes are associated with the modes of the scalar field in terms of the laws of thermodynamics. This also ensures that our analysis is physical. Further, the stability of the horizon is verified according to the changes in the black hole because the horizon is an important location, where the laws of thermodynamics are determined. As the thermodynamic variables of the black hole are also important for the dual WCFT$_2$, we demonstrate the relationship between the changes in the black hole and the energy spectrum of WCFT$_2$.

This paper is organized as follows. In Sec.\,\ref{sec2}, we briefly review the WAdS$_3$ black hole and the WAdS$_3$/WCFT$_2$ correspondence. In Sec.\,\ref{sec3}, the solution to the scattering of the massive scalar field is obtained based on its fluxes. In Sec.\,\ref{sec4}, the changes in the black hole by a scalar field are obtained and associated with the laws of thermodynamics. In Sec.\,\ref{sec5}, the stability of the horizon is tested under these changes. In Sec.\,\ref{sec6}, the changes in the black hole are well associated with the changes in the WCFT$_2$ energy spectrum. In Sec.\,\ref{sec7}, we summarize our results.

\section{WAdS$_3$ Black Hole and Dual WCFT$_2$}\label{sec2}

NMG is a ghost-free and parity-preserving theory in three-dimensional spacetime. The massive graviton of the NMG has two polarization states and spin-2 similar to the four-dimensional theory. The action of NMG with cosmological constant $\Lambda$ is given by \cite{Bergshoeff:2009hq}
\begin{align}\label{eq:NMGaction1}
S=\frac{1}{16\pi}\int d^3 x \sqrt{-g} \left(R-2\Lambda +\frac{1}{m_\text{g}^2}\left(R_{\mu\nu} R^{\mu\nu}-\frac{3}{8}R^2\right)\right),
\end{align}
where the mass of the graviton is denoted by $m_\text{g}$. $G$ is herein set to unity. We consider the spacelike stretched WAdS$_3$ black hole obtained from Eq.\,(\ref{eq:NMGaction1}). The metric is\cite{Anninos:2008fx,Donnay:2015iia}
\begin{align}\label{eq:WAdSBHmetric1}
ds^2& = -N(r)^2 dt^2+\ell^2 R(r)^2(d\phi+N^\phi(r)dt)^2 +\frac{\ell^4 dr^2}{4R(r)^2 N(r)^2},
\end{align}
where
\begin{align}
R(r)^2&= \frac{r}{4}\left(3(\nu^2-1)r+(\nu^2+3)(r_++r_-)-4\nu\sqrt{r_+r_-(\nu^2+3)}\right),\\
N(r)^2&=\frac{\ell^2(\nu^2+3)(r-r_+)(r-r_-)}{4R(r)^2},\quad N^{\phi}=\frac{2\nu r - \sqrt{r_+r_-(\nu^2+3)}}{2R(r)^2}.\nonumber
\end{align}
The negative cosmological constant is denoted by the AdS radius $\ell$, and inner and outer horizons are located at $r_-$ and $r_+$, respectively. There is a free parameter $\nu$ called the warped factor based on which the asymptotic WAdS$_3$ geometry is determined to be squashed or stretched\cite{Bengtsson:2005zj,Clement:2009gq}. For $\nu<1$, it becomes a squashed WAdS$_3$ spacetime, which has naked closed timelike curves (CTCs)\cite{Rooman:1998xf,Bengtsson:2005zj,Anninos:2008fx}; hence, we do not consider this parameter regime of factor $\nu$. When $\nu>1$, the boundary is a stretched WAdS$_3$, and the black hole is free from CTCs. Hence, we consider stretched black holes alone. Note that $\nu=1$ is an AdS$_3$ spacetime. The mass and angular momentum of the black hole are\cite{Clement:2009gq,Giribet:2012rr,Nam:2010ub}
\begin{align}\label{eq:massangularmomentum1}
M&=\frac{\nu(\nu^2+3)}{\ell (20\nu^2-3)}\left((r_+ + r_-)\nu-\sqrt{r_+ r_- (\nu^2+3)}\right),\\
J&=\frac{\nu(\nu^2+3)}{4\ell (20\nu^2-3)}\left((5\nu^2+3)r_+ r_- -2 \nu (r_+ + r_-)\sqrt{r_+ r_- (\nu^2+3)}\right).\nonumber
\end{align}
The extremal condition is satisfied at $r_-=r_+$. In terms of the mass and angular momentum,
\begin{align}\label{eq:extremalcondition2}
J \leq \frac{\ell(20\nu^2-3)}{4\nu(\nu^2+3)}{M}^2,
\end{align}
where the equality represents the extremal condition. Furthermore, the angular velocity at the outer horizon is\cite{Anninos:2008fx}
\begin{align}\label{eq:velocity1}
\Omega_+=\frac{2}{2r_+ \nu -\sqrt{r_+ r_- (\nu^2+3)}}.
\end{align}
According to the Wald formula\cite{Wald:1993nt}, the entropy of the WAdS$_3$ black hole is computed in \cite{Giribet:2012rr}. 
\begin{align}\label{eq:entropBH1}
S_{\text{BH}}&= \frac{8\pi \nu^3}{(20\nu^2-3)}\left(r_+ - \frac{1}{2\nu}\sqrt{(\nu^2+3)r_+ r_-}\right),
\end{align}
and the Hawking temperature is \cite{Giribet:2012rr}
\begin{align}\label{eq:temperature5}
T_\text{H}=\frac{(\nu^2+3)(r_+ - r_-)}{4\pi \ell(2\nu r_+ - \sqrt{(\nu^2+3)r_+ r_-})}.
\end{align}
We can also consider the timelike solutions from Eq.\,(\ref{eq:NMGaction1}), which are also divided into squashed and stretched cases. Although the timelike stretched solution includes CTCs, it is herein considered as the corresponding vacuum to a spacelike stretched WAdS$_3$ black hole. The metric of the timelike stretched solution is that of the G\"{o}del spacetime\cite{Banados:2005da}. The mass of the G\"{o}del spacetime is\cite{Donnay:2015joa}
\begin{align}
M_{\text{G}}=-\frac{4\ell^2\omega^2}{(19\ell^2\omega^2-2)},\quad \omega^2\ell^2=\frac{2\nu^2}{3-\nu^2},
\end{align}
which plays an important role as the vacuum of the WAdS$_3$ black hole. For the AdS$_3$ black hole, the diffeomorphism generators obey a specific Virasoro algebra, preserving the boundary condition. According to the central charge of the algebra, we can determine its dual CFT$_2$\cite{Strominger:1997eq}. Under the AdS$_3$/CFT$_2$ correspondence, the Bekenstein-Hawking entropy of the AdS$_3$ black hole exactly matches the Cardy formula of the dual CFT$_2$. In terms of the WAdS$_3$ black hole, the correspondence is for $\nu=1$, and a similar dual description is now found for $\nu>1$. This is the WAdS$_3$/WCFT$_2$ correspondence\cite{Donnay:2015iia}, which is briefly reviewed here.

The asymptotic boundary condition in Eq.\,(\ref{eq:WAdSBHmetric1}) is \cite{Compere:2009zj}
\begin{align}\label{eq:boundarycondition23}
g_{tt}&=\ell^2 + \mathcal{O}(r^{-1}),\quad g_{tr}=\mathcal{O}(r^{-2}),\quad g_{t\phi}=\ell^2 \nu r+\mathcal{O}(1),\\
g_{rr}&=\frac{\ell^2}{(\nu^2+3)r^2} + \mathcal{O}(r^{-3}),\quad g_{r\phi}=\mathcal{O}(r^{-1}),\quad  g_{\phi\phi}=\frac{3}{4}r^2\ell^2(\nu^2-1)+\mathcal{O}(r),\nonumber
\end{align}
which can be preserved by the diffeomorphism generators. Furthermore, these generators obey the semidirect sum of Witt algebra. Based on the generators, the algebra of the charges is constructed and found to be the semidirect sum of the Virasoro algebra generated by $L_n^\pm$. Subsequently,WCFT$_2$ is determined according to Virasoro algebra. Moreover, the eigenvalues of energy spectra $h^\pm$ with respect to $L_0^\pm$ can be associated with the charges of the WAdS$_3$ black hole. To complete the correspondence, the zero-th state of the dual WCFT$_2$ must be coincident with the zero-mass WAdS$_3$ black hole\cite{Detournay:2012pc}. This is achieved by introducing shifted operators
\begin{align}\label{eq:spectrawcft01}
h^+\rightarrow \tilde{h}^+=\frac{1}{k}{M}^2-{J},\quad h^-\rightarrow \tilde{h}^-=\frac{1}{k}{M}^2,\quad k=\frac{4\nu(\nu^2+3)}{\ell(20\nu^2-3)}.
\end{align}
Such an association between WCFT$_2$ and WAdS$_3$ exactly matches the computation of the Cardy formula and the Bekenstein-Hawking entropy. The Cardy formula is as follows\cite{Detournay:2012pc}
\begin{align}
S_{\text{CFT}}=2\pi \sqrt{-4\tilde{h}^{-\text{(vac)}}\tilde{h}^-}+2\pi\sqrt{-4\tilde{h}^{+\text{(vac)}}\tilde{h}^+},
\end{align}
where $\tilde{h}^{\pm\text{(vac)}}$ is the value of $\tilde{h}^\pm$ at the vacuum geometry. Here, the vacuum geometry is found to the G\"{o}del geometry\cite{Detournay:2012pc,Donnay:2015iia}. Then, 
\begin{align}
\tilde{h}^{\pm\text{(vac)}}=\frac{1}{k}({M}^{\text{(vac)}})^2,\quad {M}^{\text{(vac)}}=i{M}_{\text{G}}.
\end{align}
This indicates that
\begin{align}
S_\text{CFT}=S_\text{BH}.
\end{align}
The Cardy formula computed in WCFT$_2$ exactly corresponds to the Bekenstein-Hawking entropy of the WAdS$_3$ black hole. This is called the WAdS$_3$/WCFT$_2$ correspondence\cite{Donnay:2015iia}. We investigate the changes in the WAdS$_3$ black hole side due to the scatting of a scalar field, under WAdS$_3$/WCFT$_2$ correspondence.

\section{Scattering of Massive Scalar Field}\label{sec3}

We assume that an external scalar field is scattered by a WAdS$_3$ black hole. The black hole absorbs the scalar field under scattering, and changes with the absorbed amount of the scalar field. The change in the state of the black hole can be measured in terms of the fluxes representing the transferred conserved quantities of the scalar field, when the scalar field passes through the outer horizon of the black hole. The fluxes are computed from the solution of the scalar field at the outer horizon. A massive scalar field is assumed, and its action is given by
\begin{align}
S_\Phi = -\frac{1}{2} \int d^3 \sqrt{-g}(\partial_\mu \Phi \partial^\mu \Phi^*+\mu^2 \Phi \Phi^*),
\end{align}
which is a complex scalar field with mass $\mu$. Then, the field equations are
\begin{align}\label{eq:fieldeq2}
\frac{1}{\sqrt{-g}}\partial_\mu (\sqrt{-g}g^{\mu\nu} \partial_\nu \Phi)-\mu^2\Phi=0,\quad \frac{1}{\sqrt{-g}}\partial_\mu (\sqrt{-g}g^{\mu\nu} \partial_\nu \Phi^*)-\mu^2\Phi^*=0.
\end{align}
We take the ansatz to be
\begin{align}
\Phi(t,r,\phi)=e^{-i\omega t}e^{im\phi}\mathcal{R}(r),
\end{align}
where $\mathcal{R}(r)$ is the radial function to be solved. The radial equation is nontrivial in Eq.\,(\ref{eq:fieldeq2}), which is expressed as
\begin{align}\label{eq:fieldeq3}
\frac{1}{\mathcal{R}}\partial_r (\frac{4N^2 R^2}{\ell^4} \partial_r \mathcal{R})+\frac{1}{N^2} (\omega +N^\phi m )^2-\frac{m^2}{R^2 \ell^2} -\mu^2=0.
\end{align}
The radial equation in Eq.\,(\ref{eq:fieldeq3}) reduces to a Schr\"{o}dinger-type equation under the tortoise coordinate. The transformation for the tortoise coordinate is defined as 
\begin{align}
\frac{dr^*}{dr}=\frac{\ell^4}{4N^2 R^2}.
\end{align}
The Schr\"{o}dinger-type equation is obtained as follows
\begin{align}\label{eq:stypeeq1}
\frac{1}{\mathcal{R}}\partial^2_{r^*} \mathcal{R}+\frac{4R^2}{\ell^4} (\omega +N^\phi m )^2-\frac{4N^2 R^2}{\ell^4}\left(\frac{m^2}{R^2 \ell^2} +\mu^2\right)=0,
\end{align}
We consider the fluxes of the scalar field flowing into the black hole. When the scalar field configuration is within the outer horizon, it cannot be measured by a static observer. Therefore, the observable regime is limited outside the horizon. Furthermore, the scalar field inside the horizon is not distinguishable from the black hole. Accordingly, the scalar field can be assumed to be absorbed when it passes through the horizon. The magnitude of the scalar field flowing into the black hole can be measured by the fluxes of the scalar field at the horizon. Thus, we only need to solve Eq.\,(\ref{eq:stypeeq1}) at the outer horizon. In the limit of $r\rightarrow r_+$, the Schr\"{o}dinger-type equation reduces to
\begin{align}
\frac{1}{\mathcal{R}}\partial^2_{r^*} \mathcal{R}+\frac{4R_+^2}{\ell^4} (\omega -\Omega_+ m )^2=0,\quad R^2_+\equiv R^2(r_+).
\end{align}
The solutions are
\begin{align}
\mathcal{R}=e^{\pm i\left(\frac{2R_+}{\ell^2}(\omega -\Omega_+ m )\right)r^*},
\end{align}
which are ingoing and outgoing radial waves at the outer horizon. The scalar field can be only ingoing, into the black hole under the initial scattering condition. The ingoing solution and its conjugate are obtained as
\begin{align}\label{eq:stypeeq1sol2}
\Phi=e^{-i\omega t}e^{- i\left(\frac{2R_+}{\ell^2}(\omega -\Omega_+ m )\right)r^*}e^{im\phi},\quad \Phi^*=e^{i\omega t}e^{i\left(\frac{2R_+}{\ell^2}(\omega -\Omega_+ m )\right)r^*}e^{-im\phi}.
\end{align}
Note that the scalar field continues to have an outgoing case from the black hole: superradiance. This depends on the sign of the exponent in Eq.\,(\ref{eq:stypeeq1sol2}). When $\omega <\Omega_+ m $, the sign of the exponent becomes negative, and the scalar field can be emitted away from the black hole. Our analysis considers all the cases in Eq.\,(\ref{eq:stypeeq1sol2}); hence, such situations are also included in our discussion.

The fluxes of the scalar field measured at the outer horizon can be obtained from the solutions of Eq.\,(\ref{eq:stypeeq1sol2}), which are at the outer horizon. According to the fluxes, two conserved quantities, the energy and angular momentum, are carried into the black hole. The fluxes are given in terms of the energy-momentum tensor.
\begin{align}
T^\mu_\nu=\frac{1}{2}\partial^\mu \Phi \partial_\nu \Phi^* +\frac{1}{2}\partial^\mu \Phi^* \partial_\nu \Phi-\delta^\mu_\nu \left(\frac{1}{2}\partial^\mu \Phi \partial_\mu \Phi^*-\frac{1}{2}\mu^2 \Phi \Phi^*\right),
\end{align}
and the fluxes at the outer horizon can be obtained as
\begin{align}\label{eq:fluxesEandJ01}
\frac{dE}{dt}&=\int T^r_t \sqrt{-g}d\phi= 2\pi\ell R_+ \omega(\omega -\Omega_+ m ),\\
\frac{dL}{dt}&=-\int T^r_\phi \sqrt{-g}d\phi=2\pi\ell R_+ m(\omega -\Omega_+ m ).\nonumber
\end{align}
The fluxes imply that the energy and angular momentum of the scalar field flow into the black hole per unit time. Here, we assume that the energy and angular momentum of the scalar field are absorbed and transferred into the conserved quantities of the black hole as much as the fluxes. Hence, the mass and angular momentum of the black hole change the energy and angular momentum of the scalar field for a given period. Then, during an infinitesimally small time interval $dt$, the mass and angular momentum change as
\begin{align}\label{eq:changesMJ}
dM&=2\pi\ell R_+ \omega(\omega -\Omega_+ m )dt,\\
dJ&=2\pi\ell R_+ m(\omega -\Omega_+ m )dt.\nonumber
\end{align}
According to Eq.\,(\ref{eq:changesMJ}), we can the obtain changes in the initial black hole during an infinitesimal time interval $dt$. As the conserved quantities of the scalar field are considerably small compared to the black hole, we can expect the changes in the black hole to be infinitesimal. Furthermore, the leading terms of the changes in the black hole can be exactly found in the infinitesimal time interval.

\section{Thermodynamics in WAdS$_3$ Black Holes}\label{sec4}

The fluxes of the scalar field change the mass and angular momentum of the black hole, which determine its various properties. Here, we mainly focus on the changes in the thermodynamic variables and the laws of thermodynamics.

In our analysis, the mass and angular momentum increase or decrease according to the modes of the scalar field. As the locations of the inner and outer horizons depend on the mass and angular momentum in Eq.\,(\ref{eq:massangularmomentum1}), the changes in the mass and angular momentum affect the horizon location. Therefore, the initial horizons $r_\pm (M,J)$ become the final horizons $r_\pm (M+dM,J+dJ)$, where $dM$ and $dJ$ are determined by Eq.\,(\ref{eq:changesMJ}), and the changes are infinitesimally small as assumed. For the first order, the changes in the horizons are
\begin{align}\label{eq:rpmchange1}
dr_\pm&=dr_\pm(M+dM,J+dJ)-dr_\pm(M,J)= \frac{\partial r_\pm}{\partial M}dM+\frac{\partial r_\pm}{\partial J}dJ,\\
&=2\pi\ell R_+ (\omega-\Omega_+ m)\left(\omega\frac{\partial r_\pm}{\partial M}+m\frac{\partial r_\pm}{\partial J}\right)dt,\nonumber
\end{align}
where Eq.\,(\ref{eq:massangularmomentum1}) is considered, and
\begin{align}
\frac{\partial r_\pm}{\partial M}&=\frac{2r_\pm \ell (20\nu^2-3)}{\nu(\nu^2+3)(2r_\pm \nu - \sqrt{r_+ r_- (\nu^2+3)})},\\
\frac{\partial r_\pm}{\partial J}&=\frac{4r_\pm \ell (20\nu^2-3)}{\nu(\nu^2+3)(-r_\mp \nu \sqrt{r_+ r_- (\nu^2+3)}-3r_\pm\nu\sqrt{r_+ r_- (\nu^2+3)}+r_+ r_- (5\nu^2+3))}.\nonumber
\end{align}
This indicates the changes in the inner and outer horizons according to the fluxes of the scalar field, but there is no specific direction for the changes because the external scalar field can be in arbitrary modes.

Then, if the mass of the black hole decreases, there is a possibility that the Bekenstein-Hawking entropy can decrease in the final state of the black hole. For testing this, the change in the entropy is given by
\begin{align}
dS_\text{BH}=S_\text{BH}(r_++dr_+,r_-+dr_-)-S_\text{BH}(r_+,r_-)=\frac{\partial S_\text{BH}}{\partial r_+}dr_+ +\frac{\partial S_\text{BH}}{\partial r_-}dr_-,
\end{align}
and
\begin{align}
\frac{\partial S_\text{BH}}{\partial r_+}=\frac{2\pi\nu^2(4r_+\nu - \sqrt{r_+ r_- (\nu^2+3)})}{(20\nu^2-3)r_+},\quad \frac{\partial S_\text{BH}}{\partial r_-}=\frac{2\pi \nu^2 \sqrt{r_+ r_- (\nu^2+3)}}{(3-20\nu^2)r_-},
\end{align}
where we compute the first-order variation in terms of variables $r_+$ and $r_-$ with Eq.\,(\ref{eq:rpmchange1}) for simplicity, rather than using $M$ and $J$. Although Eq.\,(\ref{eq:rpmchange1}) is complicated, the entropy change is obtained in a simple form.
\begin{align}\label{eq:changeinentropy1}
dS_\text{BH}=\frac{8\pi^2 R_+ \ell^2(\omega-\Omega_+m)^2(2r_+ \nu - \sqrt{r_+ r_- (\nu^2+3)})}{(\nu^2+3)(r_+ - r_-)}>0.
\end{align}
This establishes that the entropy is irreducible under the scattering of a scalar field in arbitrary modes. Hence, the second law of thermodynamics remains valid in our analysis. Furthermore, in combination with Eqs.\,(\ref{eq:velocity1}), (\ref{eq:temperature5}), (\ref{eq:changesMJ}), and (\ref{eq:changeinentropy1}), the change in the mass is rewritten as 
\begin{align}
dM&=T_\text{H}dS_\text{BH}+\Omega_+ dJ.
\end{align}
This is exactly the first law of thermodynamics. Therefore, the first and second laws are well satisfied for arbitrary modes of the scattering of the scalar field, including the superradiance.

\begin{figure}[h]
\centering
\subfigure[{For $\nu=2$, $\omega= 1$, $m=0$.}] {\includegraphics[scale=0.51,keepaspectratio]{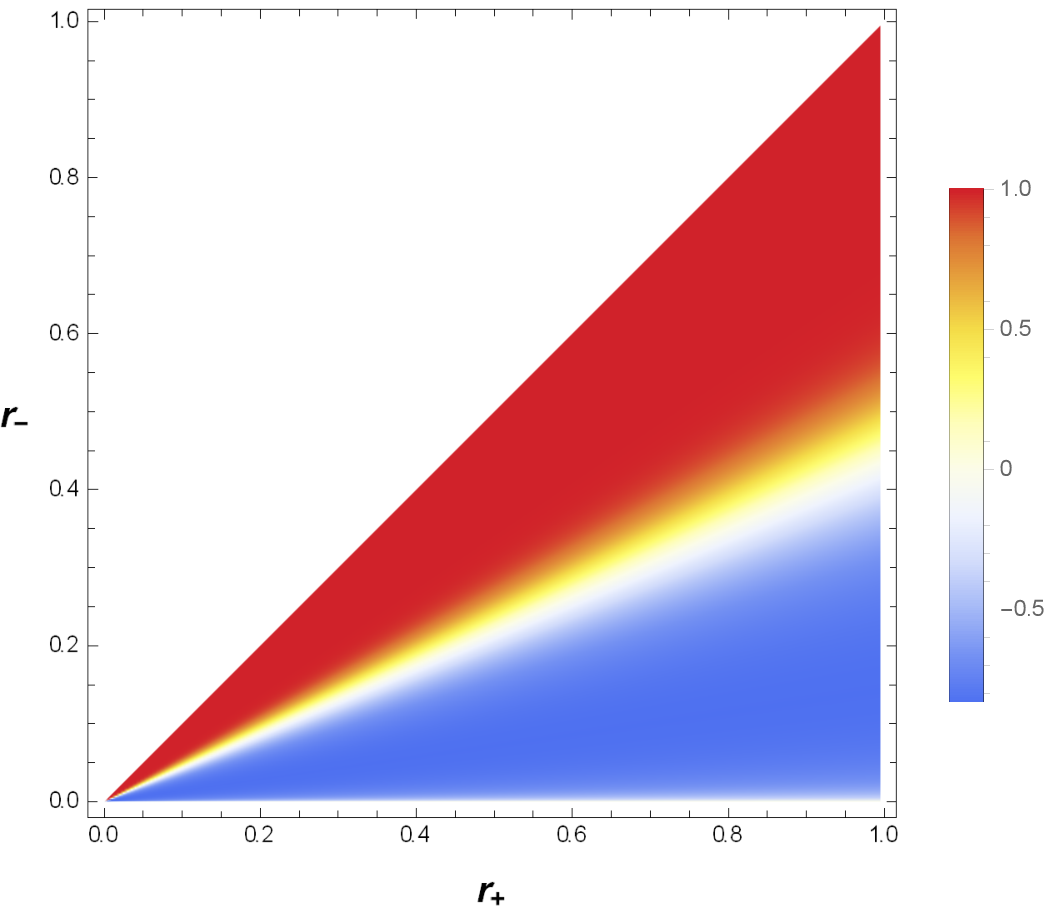}}\quad
\subfigure[{For $\nu=2$, $\omega= 2$, $m= 1$.}] {\includegraphics[scale=0.51,keepaspectratio]{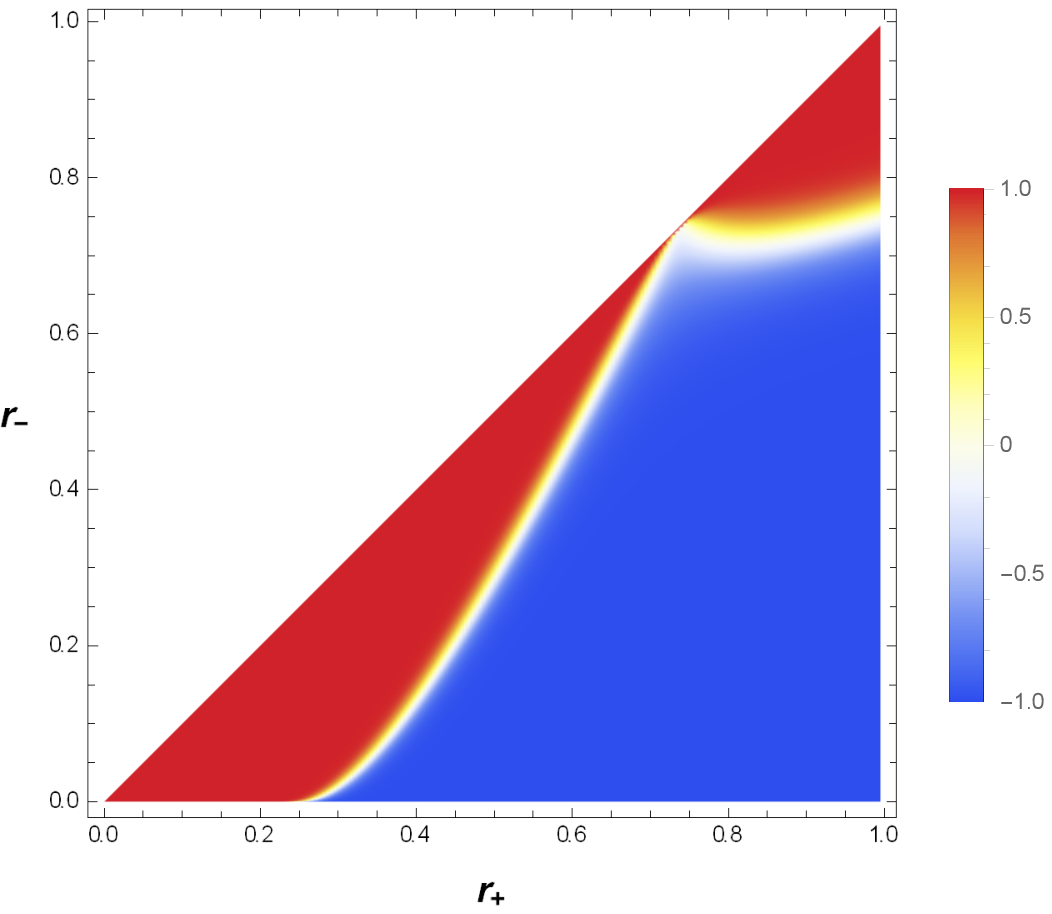}}\quad
\subfigure[{For $\nu=3$, $\omega= 2$, $m= 2$.}] {\includegraphics[scale=0.51,keepaspectratio]{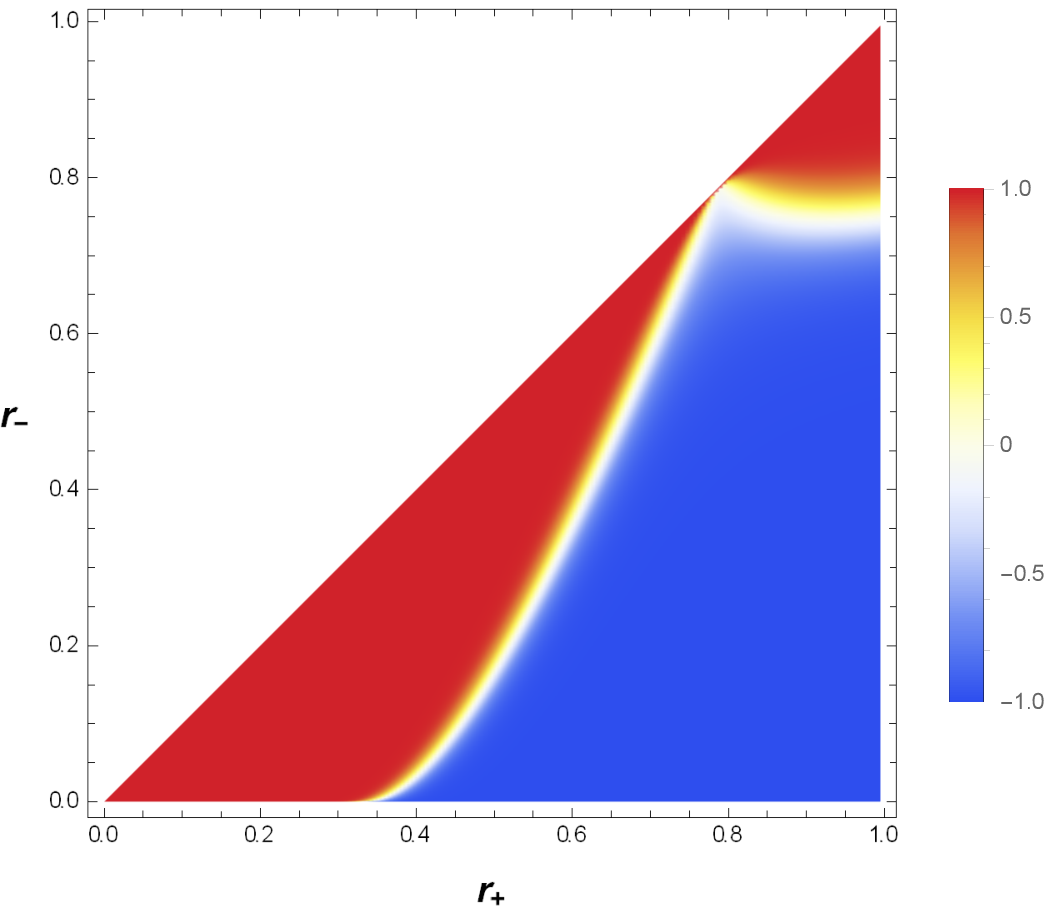}}
\caption{{\small Diagrams for $\tanh\left({dT_\text{H}}/{dt}\right)$ in WAdS$_3$ black holes with the mode $(\omega,m)$ of the massive scalar field.}}
\label{fig:fig1}
\end{figure}

The Hawking temperature becomes zero at the extremal black hole. The third law of thermodynamics is related to the physical possibility of zero temperature. Here, we consider the third law given by Bardeen, Carter, and Hawking\cite{Bardeen:1973gs}. As the surface gravity of the black hole is proportional to the temperature, we investigate whether zero temperature is possible through a finite physical process. Note that the third law is subtle compared to the first and second laws; hence, its detailed definitions or physical situations can be violated\cite{Wald:1997qp}. The change in the Hawking temperature is
\begin{align}\label{eq:changeintemperature3}
dT_\text{H}=T_\text{H}(r_++dr_+,r_-+dr_-)-T_\text{H}(r_+,r_-)=\frac{\partial T_\text{H}}{\partial r_+}dr_+ + \frac{\partial T_\text{H}}{\partial r_-}dr_-,
\end{align}
where
\begin{align}
\frac{\partial T_\text{H}}{\partial r_\pm}=\mp\frac{(\nu^2+3)(r_+ \sqrt{r_+ r_- (\nu^2+3)}+r_-(-4r_+ \nu+ \sqrt{r_+ r_- (\nu^2+3)}))}{8\pi r_\pm \ell (-2 r_+ \nu + \sqrt{r_+ r_- (\nu^2+3)})^2}.
\end{align}
As the detailed form with Eq.\,(\ref{eq:rpmchange1}) is complicated, we have omitted it here. The change in the Hawking temperature depends on the initial state of the black hole and the modes of the scalar field, as shown in Fig.\,\ref{fig:fig1} with $\ell=1$. 
The colored regime is the initial state of the black hole in terms of $r_+$ and $r_-$. Each point of the diagram represents $tanh(dT_\text{H})/dt$ for the initial state, when mode $(\omega,m)$ of the scalar field is scattered. The colors in the diagram, which correspond to the sign of the change, depict a complicated pattern. However, we can find that one pattern is coincident in all the diagrams for any parameter, wherein the change in the temperature is always positive (red color) in near-extremal black holes with $r_+=r_-$. This can be verified through an analytical approach. To expand Eq.\,(\ref{eq:changeintemperature3}) in the near-extremal black hole, we introduce an additional variable $\epsilon\ll 1$ and set $r_+ - r_- = \epsilon$. Then, the leading term in the expansion is obtained as
\begin{align}
dT_\text{H}=\frac{\ell (20\nu^2-3) R_e (\omega-\Omega_\text{e} m )^2 dt}{2\nu^3\epsilon} + \mathcal{O}(\epsilon^0),\quad R_e\equiv \left.R_+\right|_{r_+=r_-},\quad \Omega_e\equiv \left.\Omega_+\right|_{r_+=r_-}.
\end{align}
Therefore, the temperature in the near-extremal black hole increases with the scattering. The infinitesimally small changes in the black hole by the scalar field imply that zero temperature cannot be achieved by this process. Therefore, in our analysis, the third law of thermodynamics is valid. We herein ensure that our analysis under the scattering of a scalar field is valid with respect to the laws of thermodynamics. Thus, our approach to the black hole is physical, and we apply this approach to the stability of the horizon and the WAdS$_3$/WCFT$_2$ correspondence.

\section{Stability of Horizon}\label{sec5}

The outer horizon not only divides the inside and outside of the black hole, but is also the location where the thermodynamic variables are defined. The mass and angular momentum of the black hole change according to the fluxes of the scalar field. This also causes changes in the outer and inner horizons, which are functions of the mass and angular momentum. We consider whether the horizons continue exist in the final state. This is an investigation on the stability of the black hole because the horizons are significant in defining a black hole.

The locations of the outer and inner horizons are determined using function $g^{rr}$ of the metric in Eq.\,(\ref{eq:WAdSBHmetric1}). Furthermore, the locations are clearly associated with the mass and angular momentum in Eq.\,(\ref{eq:massangularmomentum1}); hence, the fluxes also change function $g^{rr}$. We investigate the change in function $g^{rr}$ and determine whether the black hole can be overspun. We assume that the initial state is a near-extremal black hole including an extremal one because the near-extremal case may be overspun by a small transferred angular momentum. This can be shown by the sign of the minimum value with respect to function $g^{rr}$. When the WAdS$_3$ black hole is near-extremal,
\begin{align}\label{eq:nearextremality03}
F_\text{min}(r_+,r_-,r_\text{min})\equiv \left.g^{rr}\right|_{r=r_\text{min}}=\left.\frac{4R(r)^2 N(r)^2}{\ell^4}\right|_{r=r_\text{min}}=-\delta,
\end{align}
and
\begin{align}\label{eq:nearextremality04}
dF_\text{min}(r_+,r_-,r_\text{min})\equiv\frac{\partial}{\partial r}\left.\left(\frac{4R(r)^2 N(r)^2}{\ell^4}\right)\right|_{r=r_\text{min}}=\frac{\partial F_\text{min}}{\partial r_\text{min}}=0,
\end{align}
where the near-extremal condition is represented by $\delta\ll 1$, and the location of the minimum is denoted by $r_\text{min}$. Note that $\delta=0$ indicates the extremal black hole. Eq.\,(\ref{eq:nearextremality03}) shows the minimum value in the initial state. As function $g^{rr}$ is minimum, the minimum value $F_\text{min}$ is small and negative. Eq.\,(\ref{eq:nearextremality04}) indicates that the location of the minimum is $r_\text{min}$. Therefore, the fluxes of the scalar field change the black hole. In the final state, the minimum can differ from that of the initial state, depending on the sign of the minimum value. When the minimum value is negative, there are no proper horizons in the spacetime; hence, the horizon is unstable, but the others imply that the horizons continue to exist in spacetime. Then,
\begin{align}
F_\text{min}(r_++dr_+,r_-+dr_-,r_\text{min}+dr_\text{min})&=F_\text{min}(r_+,r_-,r_\text{min})+\frac{\partial F_\text{min}}{\partial r_+}dr_+ +\frac{\partial F_\text{min}}{\partial r_-}dr_-+\frac{\partial F_\text{min}}{\partial r_\text{min}}dr_\text{min},\nonumber\\
&=-\delta+\frac{\partial F_\text{min}}{\partial r_+}dr_+ +\frac{\partial F_\text{min}}{\partial r_-}dr_-.
\end{align}
According to Eq.\,(\ref{eq:nearextremality03}), we can find the location of the minimum and $\delta$ by setting $\epsilon\ll 1$.
\begin{align}
r_\text{min}=\frac{r_++r_-}{2},\quad r_+-r_-=\epsilon,\quad \delta=-\frac{\epsilon^2 (\nu^2+3)}{4\ell^2}.
\end{align}
In the final state, the minimum value is obtained with respect to the first order.
\begin{align}\label{eq:minimum07}
F_\text{min}(r_++dr_+,r_-+dr_-,r_\text{min}+dr_\text{min})=-\frac{(\omega-\Omega_e m)^2 dt G \pi R_+ (20\nu^2-3)}{\nu^3}+\mathcal{O}(\epsilon),
\end{align}
where the first term is negative. This implies that the minimum value is always negative. Hence, the horizons stably exist in spacetime. Even if the scalar field changes the black hole, the change in the mass exceeds that of the angular momentum to satisfy the extremal condition in Eq.\,(\ref{eq:extremalcondition2}).
\begin{figure}[h]
\centering
\subfigure[{For $\nu=2$.}] {\includegraphics[scale=0.6,keepaspectratio]{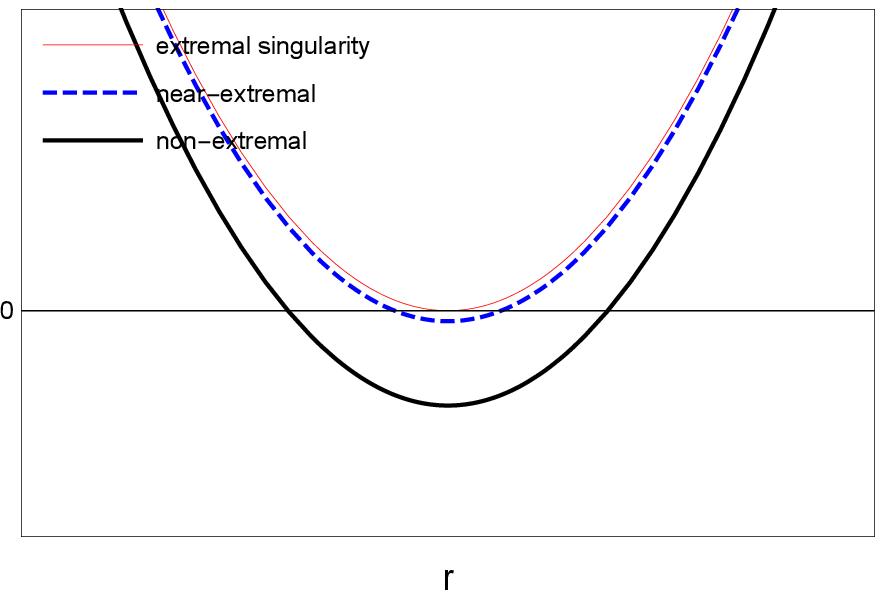}}\quad
\subfigure[{For $\nu=3$.}] {\includegraphics[scale=0.6,keepaspectratio]{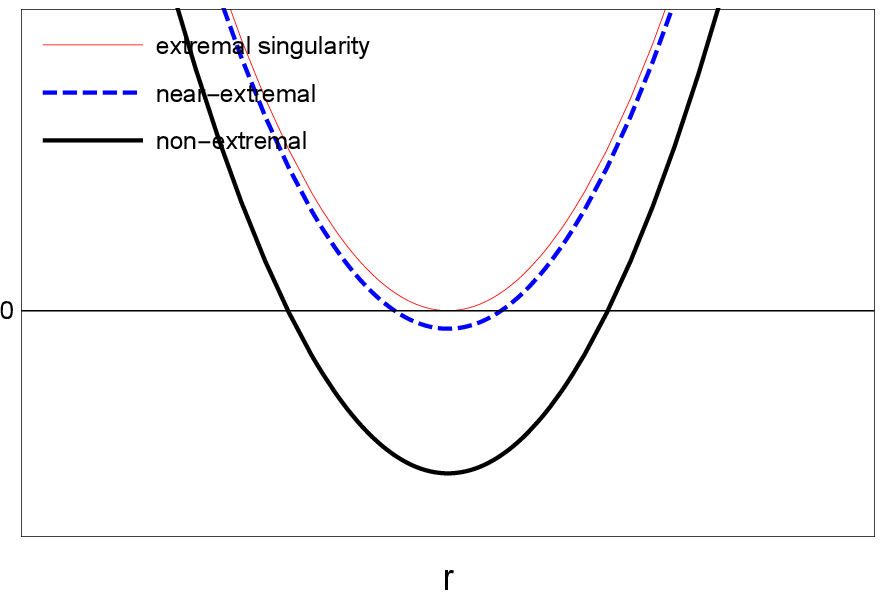}}\quad
\subfigure[{For $\nu=3.5$.}] {\includegraphics[scale=0.6,keepaspectratio]{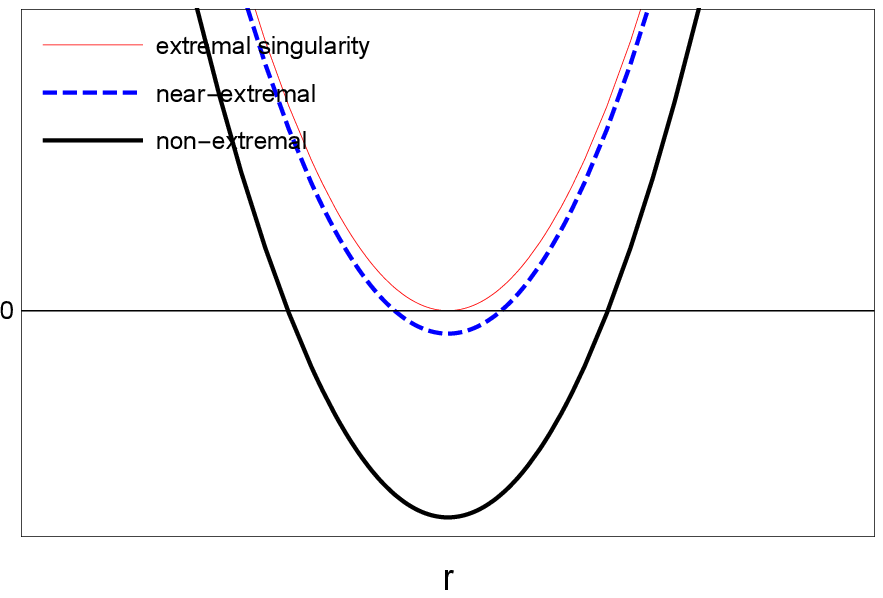}}
\caption{{\small Graphs for the function $g^{rr}$.}}
\label{fig:fig2}
\end{figure}
In Fig.\,\ref{fig:fig2}, function $g^{rr}$ is plotted for a given $\nu$. Our result demonstrates that the extremal and near-extremal states (red and blue lines) become non-extremal states (black line) because the minimum value becomes a larger negative value in Eq.\,(\ref{eq:minimum07}). In addition, we determine that the extremal black hole with $\epsilon=0$ becomes a nonextremal black hole because the first order is independent of $\epsilon$ and negative.

\section{Thermodynamics and WAdS$_3$/WCFT$_2$ Correspondence}\label{sec6}

According to the WAdS$_3$/WCFT$_2$ correspondence, the energy spectra of WCFT$_2$ are associated with the conserved quantities of the WAdS$_3$ black hole. As mentioned in the previous sections, there are physical boundaries with respect to the state of the black hole originating from the laws of thermodynamics. Then, by association, the boundaries in the WAdS$_3$ side can play a role in the WCFT$_2$ side. We investigate how the changes in the black hole are coincident with those of the shifted spectra in WCFT$_2$.

The shifted energy spectrum $\tilde{h}^+$ can be given as the function of $M$ and $J$, according to Eq.\,(\ref{eq:spectrawcft01}). When the fluxes change the black hole, the mass and angular momentum can vary as $M+dM$ and $J+dJ$, associated with a slightly different energy spectrum $\tilde{h}^++d\tilde{h}^+$. In combination with Eq.\,(\ref{eq:changesMJ}),
\begin{align}\label{eq:plusspectrum01}
\tilde{h}^+(M+dM,J+dJ)-\tilde{h}^+(M,J)=d\tilde{h}^+&=\frac{\partial \tilde{h}^+}{\partial M}dM+\frac{\partial \tilde{h}^+}{\partial J}dJ,\\
&=\frac{4\pi \ell R_+M}{k} (\omega-\Omega_+ m)(\omega-\frac{k}{2M} m)dt.\nonumber
\end{align}
This clearly depends on the modes of the scalar field. Furthermore, Eq.\,(\ref{eq:spectrawcft01}) shows that the energy spectrum $\tilde{h}^+$ is directly related to the inequality with respect to the extremal condition. This plays an important role in the change of an extremal black hole. Under the extremal condition, Eq.\,(\ref{eq:plusspectrum01}) becomes 
\begin{align}
d\tilde{h}^+=\frac{4\pi \ell R_eM}{k} (\omega-\Omega_e m)^2dt,
\end{align}
which is positive for any mode of the scalar field. Moreover, as the energy spectrum $\tilde{h}^-$ is proportional to $M^2$, its value is always positive. The change in $\tilde{h}^-$ is given by 
\begin{align}
d\tilde{h}^-=\frac{2}{k} M dM= \frac{4\pi\ell M R_+}{k} \omega (\omega-\Omega_+ m)dt. 
\end{align}
This depends on the inequality between $\omega/m$ and $\Omega_+$. When $\omega/m<\Omega_+$, the energy spectrum $\tilde{h}^-$ decreases; however, when $\omega/m>\Omega_+$, the energy spectrum $\tilde{h}^-$ increases. The condition works similar to the energy flux in Eq.\,(\ref{eq:fluxesEandJ01}). Then, according to the changes in the energy spectra, the Cardy formula changes as
\begin{align}
S_\text{CFT}(\tilde{h}^++d\tilde{h}^+,\tilde{h}^-+d\tilde{h}^-)-S_\text{CFT}(\tilde{h}^+,\tilde{h}^-)=dS_\text{CFT}=\frac{2\pi i {M}^{\text{vac}}}{\sqrt{k}} \left(\frac{d\tilde{h}^+}{\sqrt{\tilde{h}^+}}+\frac{d\tilde{h}^-}{\sqrt{\tilde{h}^-}}\right)=dS_\text{BH}.
\end{align}
Therefore, the change in the Cardy formula is exactly coincident with that of the Bekenstein-Hawking entropy in Eq.\,(\ref{eq:changeinentropy1}). This establishes that the WAdS$_3$/CFT$_2$ correspondence well works in the first-order variation.

\section{Summary}\label{sec7}

We investigated the stability of the horizons in a WAdS$_3$ black hole, which is the solution to the NMG. The horizon of a black hole is the surface on which the thermodynamic variables are defined; hence, its stability is closely related to the thermodynamics of the black hole. Particularly, in our case, the stability is also important in the WAdS$_3$/WCFT$_2$ correspondence. In this study, we mainly consider the stability of the horizons under the scattering of a scalar field. Under scattering, the conserved quantities of the scalar field can be absorbed into the black hole. Then, according to the amount of conserved quantities of the scalar field, the black hole can be perturbed. We investigated the changes in the black hole through perturbation. The amount of conserved quantities flowing into the black hole was measured based on the fluxes of the scalar field at the outer horizon. Due to the fluxes, the mass and angular momentum of the black hole change its thermodynamic properties. Interestingly, the changes in the mass and angular momentum of the black hole have a dispersion relationship, which can be exactly expressed as the first law of thermodynamics. Furthermore, the Bekenstein--Hawking entropy is irreducible under scattering. This is coincident with the second law of thermodynamics. In addition, we tested the third law of thermodynamics in this process. As the Hawking temperature becomes high in near-extremal black holes for any mode of the scalar field, the temperature cannot be zero by scattering. This is exactly ensured by the third law. The fluxes of the scalar field change all the variables, mass, and angular momentum, which determine the metric functions. The changes in the metric function imply that the horizons exist stably, and the extremal black hole becomes a non-extremal one. Then, the changes in the black hole are related to energy spectra $\tilde{h}^\pm$, according to the WAdS$_3$/WCFT$_2$ correspondence. Energy spectrum $\tilde{h}^+$ is irreducible by the changes in the side of the black hole, but energy spectrum $\tilde{h}^-$ depends on the modes of the scalar field. The change in the Cardy formula is also identified with that of the Bekenstein-Hawking entropy in the first-order variation.

\vspace{10pt} 

\noindent{\bf Acknowledgments}

\noindent This work was supported by the National Research Foundation of Korea (NRF) grant funded by the Korea government (MSIT) (NRF-2018R1C1B6004349) and the Dongguk University Research Fund of 2020.\\

\end{document}